## A heuristic sketch how it could fit all together with time

Knud Thomsen, Paul Scherrer Institut, 5232 Villigen, Switzerland, knud@confluentia-institut.ch

On a scientific meta-level, it is discussed how an overall understanding of the physical universe can be built on the basis of well-proven theories, observations, and recent experiments. In the light of almost a century of struggle to make (common) sense of Quantum Mechanics and to reconcile it with General Relativity, it is proposed to (for some time) forget about quantizing gravity or striving for one Theory of Everything or "Weltformel", which would describe the whole of reality seamlessly without any joints or suture marks. Instead of one single monolithic formalism, a three-legged compound approach is argued for. Quantum Mechanics, Relativity and Thermodynamics are proposed as the main pillars of reality, each with its well-defined realm, specific features, and clearly marked interfaces between the three of them. Not only classical reality, which is rather directly accessible to us, is then comprehensively modelled by their encompassing combination. Quantum phenomena are understood as undoubtedly lying at the bottom of classical physics and at the same time, they become "fully real" only when embedded in classical frames, i.e., preparation and measurements in time. It is then where thermodynamics steps in and provides the mediating glue as it does at interfaces towards gravity. Decoherence is understood as a smooth way of gradually transferring information and basically dumping entropy to a suitable environment. The aim of this short contribution is not to deliver novel quantitative results but rather to propose a comprehensive research program and to coarsely lay out a very roughly coherent sketch starting from the beginning of the one universe, which we inhabit. The all-embracing picture is claimed to be one of ("mutually induced") emergence.

### 1. Introduction

Since the dawn of science and certainly long before, humans wanted to know how the universe came about and how it is structured and working in a grand view, how it could fit all together. With myths and answers offered by religions evermore losing their convincing power, this remains an unsolved scientific question. The currently best available physical theories do not yet yield a unified or commonly agreed account. Here, an attempt is made to sketch an overarching picture of all of physical reality. The perspective in this short paper deliberately is a very wide one, and the lay-out of the contribution is meant to somehow reflect its message: the physical universe is identified as a braided self-referring structure, time as one pillar itself emerging only in mutual dependencies just like the other building blocks of our understanding. Physical laws as well as constituents and forces consequentially are best understood as abstractions, with clear-cut if --> then relations valid in limited contexts. Additional motivation for the chosen approach will become clearer towards the end of this tiny paper.

Arguments about how to understand and interpret Quantum Mechanics as one fundamental well-proven pillar are as old as the first conceptualizations of the mathematical framework itself. As a second foundation, Special and General Relativity, although markedly distinct from everyday direct human experience, appear somewhat easier to grasp. Given the paramount success of each theory in its field of application, usually many orders of magnitude apart, the very best minds have toiled for more than 100 years to somehow reconcile quantum mechanics with relativity, alas hitherto without resounding success. Thermodynamics provides another cornerstone. In its essence it is more an abstract and, in a sense, eternal mathematical theory. According to the frequently cited assessment by Albert Einstein, thermodynamics uniquely constitutes one foundation of our understanding of the world, which will within the framework of the applicability of its basic concepts not be overthrown by new findings.

Here, it is humbly suggested to take a big step back and assume a very broad perspective for a fresh look, keeping for a start to a very coarse and superficial level while exploring what a rather sketchy but all-embracing picture comprising quantum mechanics, classical physics, relativity, and thermodynamics could look like while staying firmly anchored in solid science.

With more or less obvious paths leading to no true breakthrough in unifying our currently best understanding of the universe, some apparently outlandish inspirations seem appropriate. In line with a recent account of cognition, it shall be tried to sketch a big schematic picture before details are worked out in necessary subsequent steps [1]. There is no doubt that at the end well-matching quantitative accounts, "laws", are the goal. It should just be avoided that these (the difficulties finding them) block or blur an overall view and thus hamper progress from the onset.

The aim of this paper is to assess a possible truly overarching match between currently available experimental/observational evidence and fundamental theoretical findings; it is structured in the following way: chapter 2 looks back at the very roots of interpreting quantum mechanics and argues for a well-defined Heisenberg cut. On that basis, chapters 3 and 4 investigate causality and the measurement of time with real clocks and its progression in one sense only (the arrow of time). Subsequently baselining time as emergent, it is suggested in chapter 5 that a variable "progression of time", understood as a change in the ratio between different clocks, might resolve some of the problems which had provoked the conjecture of an initial inflationary phase. The focus is further shifted to the character and role of gravity in chapters 6 and 7. In the light of the previous sections it is then argued in chapters 8 and 9 that neither dark matter or dark energy nor many worlds appear required for a consistent overall picture of the physical universe. Principal limits to achieving any full understanding are dealt with in chapter 10, while the conclusions in chapter 11 try to outline an overall self-reflective and -consistent narrative and propose some immediate upshots of the presented considerations.

## 2. Timeless Quantum World

From the very beginning of quantum physics, Niels Bohr emphasized the importance of the involved classical experimental apparatus. Quantum effects always are described and observed in a context in the real world [2]. On the coarsest "outside" level, the relevant context is classical; it frames free and undisturbed quantum states at a beginning event as well as at an endpoint. The latter ones we are usually well aware of when we record and report them. Definite measurements yield irreversible classical results, but any selected quantum state also needs some preparation harnessing classical boundary conditions [3,4]. Our everyday world is profoundly classical, and we know of ("weird") quantum phenomena only from more detached constellations and deliberately arranged experiments. Exotic behaviors of quantum systems like entanglement and superposition of states are very fragile and occur almost exclusively in carefully controlled closed set-ups and for limited time intervals, well-shielded from any disturbing influence from an outside environment [5]. It can be shown that relying fully on quantum systems also for reference, i.e., quantum reference frames (QRFs), superposition and entanglement are relational and in fact just different sides of the same coin [6,7]. Inside, quantum theory does not distinguish between prediction and retrodiction, the Born rule applies equally well in both directions of time [4].

In Wigner's friend scenarios, Wigner's measurement in general changes the friend's memory record of her result; any awareness by the friend of this change would conflict with the no-signaling condition [8]. Interfaces between quantum and classical phenomena are marked with time stamps, i.e., in most cases some type of persistent classical memory. This need not be a record set in stone, fleeting ("pale") traces suffice [9]. After some while, probably nothing more than a minute increase in overall entropy is left while some free energy has been "consumed". Importantly, at every irreversible transition (like erasure, i.e., when the input state cannot be fully recovered from an output state) there is a concomitant change, i.e., increment, in entropy in some environment outside of the narrow (quantum) system under investigation [10,11,12]. Entropy characterizes uncertainty and ignorance. After an event with a growth in entropy, it is not possible to reconstruct what had been the exact and complete state before.

It has recently been proposed that a clear Heisenberg cut can be identified with an increase of entropy and the associated transfer of energy (or "consuming" another conserved quantity) between the investigated system and its environment according to Landauer's principle [13]:

$$\Delta E \geq kT \ln 2$$

Unitary quantum systems inside their classical demarcations could then be conceptualized as "timeless". This condition provides all the internal opportunities for the exploration of each and every development, which is permitted by the boundary conditions, in parallel. All available options are taken into account as with Richard Feynman's path integrals [14]. As long as nothing irreversible occurs, all possible states will somehow potentially/ghostly co-"exist" inside an isolated QM system according to the Schrödinger equation "at the same no-time", with their respective probabilities for being observed given by the Born rule.

For the arguments here, it can be disregarded that there is in fact a whole "zoo" of different but related definitions of entropy in different (quantum) contexts [15,16,17,18]; basic Boltzmann entropy suffices for a start.

It can convincingly be argued that there is no way getting time from no-time in quantum mechanics without first presupposing some concept of time [19]. The Hamiltonian operator, corresponding to the total energy of a system, is firmly rooted in the classical reality, and it generates the time evolution of quantum states in the Schrödinger picture, in fact by defining boundary conditions possibly varying with respect to laboratory-time. For observables, speed limits apply [20]. Dynamical quantum systems are thus tightly tied to laboratory "standard" time during unitary evolution, while internally enjoying all the freedom also in their phasing such that only statistical predictions are possible before a (collapsing) measurement. In case external classical boundary conditions vary with time explicitly, the last version (defined by the Hamiltonian and by the associated classical time stamp) is decisive. Unintended disturbances interrupt the unitary evolution the same as for any static system as soon as enough energy is transferred / entropy produced.

Upon the collapse of the wavefunction, linearity and unitarity are broken, one classical outcome is materialized depending on the overall set-up, the apparatus; e.g., wave- or particle-behavior is observed [21]. This holds for single individual cases ideally. The Landauer threshold is somewhat paradoxically strictly applicable in this simple form only in quasi-static conditions; the picture is more complicated when unavoidable fluctuations are taken into account but Landauer's principle is also valid then [12,22]. Two qualifications should be added. Strictly, the bound is an entropy, k ln2 per bit; kT ln2 is the heat corresponding to it at temperature T. And the bound holds for a complete cycle, not for each single measurement: a result can be copied reversibly into a memory prepared blank, and the price has to be paid only when this memory is reset [11]; how the cost is divided between measuring and erasing depends on the memory used [23]. If the measured system is still accessible, a record can even be reversed without any cost; quantum erasure experiments exploit exactly this.

With multiple systems / repetitions, statistics can be compiled, and in an intricate set-up a quantitative complementarity relation for wave-particle behaviors has been measured [24]. It is hypothesized that weak measurements employing an ancilla disturb / collapse the system of interest only partly and for each data point effectively transfer just a tiny amount of entropy / energy; no sudden collapse ensues but entropy will accumulate gradually. Tsallis entropy might be an apt tool to describe this [18]. Dilution of entanglement in the environment, i.e., decoherence, leads basically to the same classical behavior of a considered system. A trade-off between information gain, reversibility, and disturbance has been demonstrated for quantum measurements employing ensembles [25]. The more information is extracted, the more a state is disturbed and/or the less recoverable it is, this all for a constant space frame. There is a minimum amount of entropy production required for obtaining information about work done to a quantum system driven far from equilibrium [26]. Entropic uncertainty relations can be shown to be equivalent to wave-particle duality [27,28].

Contextuality denotes the fact that measurements of quantum observables can in no case be simply thought of as revealing pre-existing values [29,30]. Results depend on which other observables are measured together in a sequence. Projective measurements of incompatible observables are not commutative; they in turn yield (new) classical results and constraints, their order matters. Surplus weirdness can be avoided when carefully keeping to the respectively applicable contexts [31,13,32].

Normally, we do not exactly know the complete frame. A single photon without a heralding companion does not tell whether or how there exist(s) any entangled state(s): no local measurement can distinguish one partner of an entangled pair from a member of a classical mixture, which is also why entanglement cannot be used for signalling.

It appears not immediately clear that at the start of an (internally) timeless phase a similar discrete event has to happen for properly conditionalizing a thereafter isolated individual quantum system. The necessity of carefully preparing a quantum state has been described by Niels Bohr, Willis Eugene Lamb, John Archibald Wheeler, Andrei Khrennikov, and many others. Given that quantum theory is time-symmetric, it only seems natural that the beginning and the end of a completely isolated quantum system are considered equivalent (except for their embedding in outside (laboratory) time).

Current conceptualizations of Bell measurements emphasize their contextuality-aspects [33,34,35]. Experiments with photons from a distant star or quasar might raise some doubt but fit nicely with the idea that the undisturbed travelling photons are somehow “out of time” and fully enter reality only upon registration. Light quanta are not detected as arriving in earmarked entangled pairs; such “fossil” light can effectively be used as a random generator for closing the freedom-of-choice loophole in Bell experiments [36]. The latter only if one does not believe in principally untestable metaphysical superdeterminism, which claims that about everything is somehow correlated or determined since the very beginning of the universe.

## 3. Clocks in the real world

The classical frames of quantum systems with their start- and end-points can, depending on the available details, be ordered in at least one way to form consistent histories. Time thus starts out locally and discrete. Events with their records have been described as creating empirical (space-)time before [37,38]. Similar ideas have been put forward elsewhere: in causal set theory, time passes as new atoms of spacetime are born [39]; in entropic dynamics, time keeps account of the change that has accumulated [40]; and an entropic time has been proposed whose moments follow one another in the order of the information gained through decoherence [41].

Not only collapse-events, but, of course, also interactions in the classical domain with energy transfer (and entropy production) generate records and timestamps. Given some energy/time uncertainty relation, “moments” come with a fuzzy extension, i.e., minimum duration. The sheer number of related and partly overlapping and also nested/embedded systems following the same laws and producing records, allows for their qualified ordering and, depending on the context, also synchronization; practically quantitative, smooth, and continuous time emerges. Employing a model of a relaxation process, time can be shown to appear as a coarse-grained parameter in the statistics of measurements of events very similar to temperature [42]. The analogy with temperature only goes so far, however: thermometers come to agree by exchanging heat, whereas clocks agree without having exchanged anything at all. Optical clocks based on different atoms, which are not coupled to each other, show frequency ratios that have been determined to 6 to 8 × $10^{-18}$ [43], and the ratio of two optical transitions in $Yb^+$ has stayed constant over years [44]. A synchronization, like every comparison, can settle conventions, a common starting point and common units, once and for all; but nothing acts on independent clocks between two comparisons, and so it cannot be the reason why they still agree at the next one. That good clocks keep step is therefore a law of its own, a zeroth law of time. It applies to classical clocks: when clocks interact via gravity, time dilation entangles them, and a single clock then has no well-defined time [45].

Any clock produces entropy, and clocks need energy, the better they are, the more. The laws of thermodynamics dictate a trade-off between the amount of heat dissipated (entropy produced) on the one hand side and the accuracy and resolution on the other [46,47,48]. This holds, however, only where the regularity of a clock stems from its ticks themselves. A model pendulum clock escapes the trade-off [49], and for a coherent quantum clock the precision can even increase exponentially with the entropy produced per tick [50]. There, just as in atomic clocks, the regularity is provided by a coherent flywheel, which writes no record in between two readings; it is the reading and the correcting that cost entropy, and none of them works without producing entropy [51].

Boundary conditions are given by relativity; time stamps are relative in a context and they are local. The dependence of outcomes on the order of such events establishes one direction of time (in a context/history), which cannot simply be reversed. This is ascribed to the entropy generated when establishing the records. These traces together with the mechanism generating them effectively constitute clocks, which cannot exist as such inside an isolated timeless quantum system. Any clock needs something which changes, either periodically or as in some form of relaxation. The expansion of the universe is a special and most important example of the latter: it provides a common cosmic time and, with it, a global direction of time [52], to which all other clocks are one way or the other related. It does not set their rates, however; bound systems such as atoms, planetary systems and galaxies do not take part in the expansion, and their clocks tick at rates fixed by local physics and by the gravitational potential they are in.

The well-known Page Wootters mechanism for supplying clocks to a quantum system involves a second system, in which memories/records are conserved (i.e., effectively classical in disguise) [53]. Indeed, if the clock in the Page Wootters mechanism is classical, the time parameter of the Schrödinger equation is recovered [54]. Even disregarding a full collapse of a quantum system, energy cannot be measured arbitrarily fast by an external system, and the evolution observed by an internal clock cannot be unitary during an energy measurement regardless whether an internal or an external system carries out the measurement [55].

In general, quantum measurements cannot be performed instantaneously in decoherence-focused interpretations; the time required scales with the change of entropy of the measured system [56]. There are speed limits on observables both classically and for quantum systems [20]. Also in the light of what has been shown for (continuous) position, it seems natural to state that time principally cannot be measured to arbitrary precision [57]. Measurements have a minimum energy cost, infinitely exact measurements quite generally would require unbounded resources [48,58].

No clock is a clock without memory [13]. Clocks with some permanence require some entropy production; if they were all unitary and reversible there would be the risk that they run in the wrong direction [46]. A definite thermodynamic time's arrow is restored by a quantum measurement of entropy production if the value found is large; for values of order k the two processes can still interfere [59]. In experiments purportedly putting counter-running arrows of time in a superposition, a quantum evolution and its time reverse were superposed [60,61]; only unitary operations were involved, and no record with entropy was generated during the process before the very (classical) end.

For periodic clocks, progressive counting is essential, and for aperiodic ones it is required to remember at least some starting value. Real physical clocks cannot be in equilibrium and they cannot be reversible, they require some reservoir of low entropy / some source of free energy. In suitably large enough real classical systems, memories, at least traces, are possible, and entropy is never decreasing; reversibility is barred, and causality can be relied on [62,63].

Building on the embedding of quantum processes in classical spacetime, it can be shown that realizations of processes with indefinite causal order (ICO) that respect relativistic causality are forbidden unless input and output agent systems are non-localized [64,65,66]. Even then, one can "zoom in" and unveil a finer level of description exhibiting a well-defined and acyclic causal order. ICO processes, which violate causal order, cannot be realized faithfully in classical spacetime.

Causality is a time-oriented abstraction on the basis of (interaction) events, in particular their observed order, and correlations relatively high up in a hierarchy of concepts following an ontology as devised by Nicolai Hartmann [67]. Human cognition unfolds in time, and it works with expectations but not in the sense that we only dream up and thus constitute regularities but rather that these are (often statistically) extracted from learning, in part already ground-laid by evolution over eons, as sketched in the Ouroboros Model [1]. Linking two events causally demands much more connection(s) between these two than mere succession in time, i.e., generally the impact of an event or intervention has to be described [38,68]. To understand anything acceptably, detailed schemata have to be activated and filled without leaving big gaps between any purported cause and effect. This in turn does not mean that clocks would require some observer to read them or that cause and effect would wait for anyone to disentangle them.

Here it is argued that one has to forego any tacit assumption that normal classical laboratory time can be directly extrapolated to arbitrary settings, e.g., like the very beginning of the universe.

## 4. The Arrow of Time

Ordered records with reversible transitions in forward and in backward direction do not suffice for keeping time, nor for establishing a well-defined and univocal direction of time passing. A most simple example is given by an old film roll, which (given suitable content) can be seen with one or the other succession of frames without noticing any uncertainty or error. Similarly, it is currently agreed knowledge that the standard microscopic laws of physics, which describe possible developments in forward direction do this equally well in backward direction [59], apart from a small violation of time-reversal symmetry in the weak interaction [69], which is not thought to cause the thermodynamic arrow.

Common agreement can be refined and superseded as a result of more information (from additional sources like, e.g., more powerful instruments) becoming available, which can yield a new and improved understanding. Recently it has been argued that, indeed, there is a microscopic classical phenomenon, which specifies an allowed forward direction of time. An accelerating wave equation is reported to have a solution only with time progressing but not in reverse [70,71]. In this approach, the long-standing Abraham-Minkowski controversy about the momentum of light in a medium is ascribed to relativistic effects [71], and a heuristic resolution by carefully considering the used frames of reference has been proposed [70]; more generally, the controversy is regarded as resolved since the two forms were identified as the kinetic and the canonical momentum of light [72]. In terms of longitudinally accelerating waves, thus there is a well-defined direction of time, an "arrow of time." Relativistic (observer dependent) effects ensure the conservation of momentum of the wave between different media. The proper time of the accelerating wave is universal and analogous to the proper time in Special and General Relativity, not necessarily the same as laboratory time. With a constant reference velocity, momentum and energy are conserved for a wave moving along a geodesic.

Time dilatation as described by Special Relativity ensures a smooth interface to the timeless quantum realm: in a system speeding up and approaching the speed of light (thus also becoming ever more isolated), internal time drags on ever more slowly, and its passing diminishes, coming to a standstill in the limit (for an outside observer). In a recent paper it has been shown that even tachyons, by definition travelling with a speed larger than the speed of light, can consistently be described by an explicitly covariant quantum field theory, just that the Hilbert space has to be doubled [73]. Taking the initial and final state of the system on equal footing as boundary conditions for the calculation of probabilities involving tachyons turns out to be necessary for relativistic invariance. It is highlighted by the authors that this singles out the two-state formalism of quantum mechanics featuring time symmetry of quantum measurement processes proposed 60 years ago [74]. A time-symmetric theory then obtains the arrow of time by macroscopic factors like an embedding in the expanding universe.

Since its first conception by Arthur Stanley Eddington, the arrow of time, which we experience in the macroscopic world, has been traced to a state of very low entropy at a beginning of the universe. Its continuous expansion goes hand in hand and delivers a backdrop with time progressing in only the forward direction [52]. With progressing time, the universe develops towards some equilibrium, and entropy is increasing. This might go on until the universe has so much expanded that almost any finite region is empty; this would then be an effectively timeless condition: in equilibrium time loses its meaning [42], and no records can be made. The entropy would not return to zero, though: by the second law it stays at its maximum, while the entropy density goes to zero. In an accelerating universe this maximum is set by the entropy of the cosmic event horizon, today about $10^{122}$ k, far above the about $10^{104}$ k inside the observable universe, most of it in supermassive black holes [75].

While there is little disagreement about the principal existence of a postulated highly ordered state with low entropy in the early universe, it is not so clear how this purportedly very special state could ever have arisen. As to time then, there is no outside observer, and a homogeneous concentration of mass causes no time dilation between places: all comoving clocks show the same cosmic time. For later observers, however, early processes do appear slowed down, by the cosmological time dilation factor 1+z, which has been measured in the light curves of distant supernovae [76]; this is a property of the light paths, not of the clocks at the time. Leaving out for now the very first moments, there is an obvious way to pinpoint an important transition in time at decoupling when the universe was about 370 000 years old and approximately 3000 K hot with expansion going on thereafter. This is documented in the cosmic microwave background, CMB [77].

The proposal is, again, to consider the full frame of reference, in particular its change from one condition to another in the early expanding universe. What was an average high entropy state of matter at an early point with basically only short-range forces, became one of very low entropy triggered by the changing balance between the contributions of the dominating forces. Gravity at that time of continued expansion becomes the most effective force on larger scales, and it is attractive. As gravity tends to clump matter together, the same homogeneous smooth state, which had high entropy as long as only short-range forces mattered, becomes a highly improbable state of low entropy compared with the clumped states that gravity makes accessible [78]. Roger Penrose with his Conformal Cyclic Cosmology, for example, has raised similar arguments (in order to bypass questions relating to special initial conditions, hypothesizing an eternal recurrent universe) [79].

Deemphasizing the role of gravity, another special point in time has been proposed by Carlo Rovelli [9]. At about "one second" after Big Bang, protons and neutrons were no longer in thermal equilibrium due to the expansion of the universe. At the freeze out temperature of about 0.7 MeV the ratio between neutrons and protons was fixed, and with it, after some neutron decay, the ratio between helium and hydrogen formed a few minutes later. Giant clouds of hydrogen can later be identified as providing one suitable reservoir of low entropy [80].

Very recent calculations arrive at the conclusion that the universe at its current state with low entropy and a small cosmological constant may actually not be so special anyway [81]. Even if coarse graining is required for defining entropy, this does not necessarily mean that the arrow of time turns fully perspectival, and could only be rescued with some type of anthropic argument.

The Bekenstein bound specifies an upper limit how much entropy can be contained in a volume with a given energy (on the surface) following the second law of thermodynamics [82]. A small baby universe would have its limits. As soon as there is an environment to which enough entropy can be dumped, quantum states can be framed (classically), and matter can turn into real, the expansion process at least then becomes irreversible. It has been argued that changes of time dilation, both from Special Relativity and from gravitational causes, can produce entropy: a thermal quantum system that changes its speed or height is driven out of equilibrium [83]. The constituents of the early universe are not accelerated by the expansion, however: comoving matter falls freely and shares one cosmic time, so that this mechanism could apply to changes of their peculiar motions, not to the expansion as such.

Each single photon of the CMB detected now has lost most of its energy since it had been emitted, redshifted due to the expansion of the universe. Propagating all the time with the speed of light, these photons cannot really be called "tired" but they are definitely "stretched". One might hypothesize that the "lost" energy went into entropy production, again adhering to Landauer's principle. This is not the case: free-streaming radiation cools adiabatically, its photon number and entropy per comoving volume stay constant while its temperature falls as $1/a$ [84], and the energy decrease is the work done by the radiation pressure in the expansion; globally, energy is not conserved in an expanding spacetime. The redshift of the CMB thus produces no records. The ratio between entropy and energy of the radiation nevertheless grows during expansion, because the energy falls while the entropy stays constant; this fits with a corresponding increase of the Bekenstein bound on that ratio [82], $2\pi kR/\hbar c$, which grows with the radius of a comoving volume as $a$, so that the radiation stays as far below the bound as before. Entropy is produced, and records are made, where radiation is absorbed and re-emitted far from equilibrium, for instance when starlight is absorbed by dust and re-radiated in the infrared.

The expanding universe in total can be seen as an "absorber", allowing only outward spreading electromagnetic waves. The microscopic direction of time from the accelerating wave equation thus fits nicely with the general cosmological arrow of time, and so do all other such arrows, e.g., our perceptual and psychological ones, too. Animals including humans actually are (inter alia) clocks employing in fact a plenitude of mechanisms in parallel, which might be one reason why a timeless quantum world is so hard to imagine and is felt to be so weird. Interestingly, Large Language Models, which are the current most successful models of human thought are slightly better at predicting what comes next in a sentence than what came before [85].

## 5. No need for inflation

In standard scenarios of Big Bang, time and gravity, and actually all of the world, starts in a state at least close to a mathematical infinity. Inflation during a first short period then is purported to finally yield a universe, which matches with current observations.

The classical preparation of an isolated quantum systems poses a challenge in any laboratory. With gravity governing the expansion, but probably assuming no decisive role in the interactions between particles long before freeze out or recombination and transparency, quantum effects of the other three fundamental forces of nature dominated. The very beginnings of the universe certainly were not classical, and the conditions of systems separation, thermodynamic imbalance and long thermalization times for memory and traces very probably were not met [9]. Far from equilibrium with time only emerging in the process and diverse non-linear feedback between all constituents, certainly no nice linear scale for the development of anything can be expected.

The proposal here is to consider the possibility that time in some way progressed from a very beginning but clocks ticked differently before the advent of hadrons and/or neutral atoms, which could provide some basis for defined (classical) boundary conditions and a suitable environment for records and for dumping entropy. The idea is trivial: something can seem extremely fast in case the available clock runs very slow. Not some intrinsic speed but rather the applicable frame and scale would be to investigate for answering open questions, e.g., concerning smoothness. A "slow genesis" of space and time during that initial phase might in hindsight just look like "inflation" (with its "timely duration" constrained also by some sort of time/energy uncertainty relation). Two qualifications are needed. If all clocks ran slowly together, nothing observable would change: a common slowing of all clocks, even a time-dependent one, only relabels time. The horizon problem concerns light cones, i.e., the comoving distance light can travel before recombination, the integral of c dt/a, and a relabelling of t leaves this integral unchanged. Measured by atomic clocks, with c constant, enlarging it requires an early phase in which the comoving Hubble radius does not grow, as in accelerated expansion, which is inflation in its kinematic sense, or in a contraction before a bounce. The idea acquires physical content as a change in the ratio between different clocks, in the sense of the "speed of time" of chapter 7: between light and expansion, as in cosmologies with a varying speed of light, which have been proposed to solve the horizon problem [86,87], or between atomic physics and expansion, as discussed next.

Presumably the Hubble tension could be addressed if atomic clocks and the expansion kept a different ratio in very early and later phases of the unfolding expanding universe, similar in effect to changing coupling constants: a slightly different electron mass at recombination has been shown to reduce the tension from about 5σ to between 3.4σ and 1.0σ, depending on model and data [88]. Probably this is not really needed, and identifying some other biases in measurements can resolve the currently observed discrepancies between diverse methods [89,90]; the latest consensus of distance-ladder measurements, however, finds the tension persisting [91]. At the time the cosmic microwave background froze out the transition most likely was not razor-sharp, and remainders of earlier "slower time" might have still been effective (the universe appears to be expanding faster in our (local and temporal) vicinity). This might just match with (almost) no time passing before decoupling (given ample opportunities for all types of development like in small isolated quantum systems), but only in the sense of a common relabelling of all clocks. A slowing of nuclear and weak-interaction clocks relative to the expansion would have changed the abundances of the light elements, and nucleosynthesis confines the expansion rate at that epoch to within a few per cent of its standard value [92]. At least from nucleosynthesis on, and demonstrably at decoupling, the expanding universe had no centre and no periphery: apart from the dipole due to our own motion, the CMB is isotropic to about $10^{-5}$, and we do not sit at a special place [93]. Close to the beginning this need not have been so. Generic solutions of General Relativity approach the singularity anisotropically and inhomogeneously [94]; the homogeneity and isotropy we observe may have emerged, for instance during inflation [95], which drives anisotropic models towards isotropy [96], or they were part of a very special initial state [97]. Time passing in denser regions would still have been gravitationally dilated compared to less dense ones, and differences in the local expansion could have had an influence, too (probably even a partly compensating one, see chapter 6); but at decoupling the differences of the gravitational potential were only of order $\Phi/c^2 \sim 10^{-5}$, and their imprint on the CMB is the Sachs–Wolfe effect [98]. Larger gradients in the rate of time would have to be reconciled with the precisely measured acoustic peaks of the CMB [99].

Inflation has been proposed to solve a number of problems like the observed homogeneity of the visible universe in every direction [95,100]. After tremendous initial success, foremost earlier proponents have turned fierce sceptics. Discrepancies between ever more exact measurements have surfaced, and it seems that in order to avoid extreme finetuning for certain parameters others have to lie in unbelievable narrow regions [79,101]. Others conclude that inflation is on a stronger footing than ever [102].

A deep arrow of time pointing in the usual direction but with "less speed" might obviate the need for an inflationary phase. Homogeneity does not always require direct interaction. Behavior of the constituents according to the same laws, which became effective after wider separation, could have led to very similar outcomes overall. Common laws alone do not make common initial conditions, however, and the near-uniform temperature of the CMB requires the latter. With one common timescale for all phenomena ascribed to the first moments after the Big Bang, it can be speculated, that changing just that scaling has only moderate impact on specific mechanisms unfolding. With space expanding, unitarity breaks down even without disturbance, and only isometry appears to be left [103,104].

## 6. Matter and antimatter

While gravity might not be of top importance directly for the progressing of time in the earliest phases it might be interesting to note that within the initial gigantic concentration of energy/mass (even if not infinite), conditions will vary to some extent. This is hardly seen today, one main reason, why an inflationary period was invented originally.

Gravity defines geometry. At least from nucleosynthesis on, the baby universe has no middle and no periphery, but it has small density fluctuations, and with them differences of the gravitational potential; some type of gradients thus appears unavoidable. Even in a constant gravity field, the temperature in a gas in thermal equilibrium and in gravitational equilibrium cannot be uniform [105]. In this case, still, a uniform temperature is seen by an observer due to gravitational redshift. Given the universality of free fall, the validity of the Stefan–Boltzmann law is not affected by a temperature gradient stemming from gravity [106]. In the CMB the corresponding compensation is incomplete: radiation from a potential well is intrinsically hotter but loses more energy climbing out, and for adiabatic fluctuations on large scales the net effect is $\Delta T/T = \Phi/3c^2$, with $\Phi$ the Newtonian potential, so that potential wells appear as cold spots [98,107]. This Sachs–Wolfe effect dominates the CMB anisotropy on large angular scales.

If gravity affected matter and antimatter slightly differently this might be the reason for the absence of antimatter in the observable universe. First measurements of the behavior of antimatter in the weak gravitational field of the earth show no difference to ordinary matter but need not be of real relevance for very high concentrations of energy and gravity. A recent experiment at CERN most probably was far too insensitive to see any difference in gravitation for matter and antimatter at a level likely required to explain the imbalance observed in the universe today [108]. Actually, it is not so obvious how any finding in such an experiment would relate to a minuscule asymmetry during baryogenesis [109].

Some tiny violation of the CPT symmetry has been proposed as one explanation of the dominating prevalence of matter over antimatter [110,111]. The standard route needs no CPT violation, but Sakharov's three conditions: violation of baryon number, violation of C and CP, and a departure from thermal equilibrium [112]. Charge and parity are quantized, but the corresponding symmetries are not solid: both are violated by the weak interaction, CP is violated too, though only weakly, and searches for further CP violation continue, e.g., for an electric dipole moment of the neutron [113]; T violation has been observed directly [69]. Within the Standard Model, however, both the CP violation and the departure from equilibrium are too small [114]. Here it is suggested that, in fact, the third condition is the crucial component: time-symmetry, in the thermodynamic sense, is violated in the relevant context far away from equilibrium. With time itself only emerging during the violent processes of expansion and baryogenesis, it certainly cannot in this epoch be considered as an effectively "static container" in which any development can run unaffectedly in forward or backward direction equal in all detail. To a much lesser extent this applies in general: already Heraclitus knew that no one can step into the same river two times.

## 7. Gravity from thermodynamics and the other way round

Gravity in turn might also be seen emergent as an entropic force. Ted Jacobson has shown that the Einstein field equations, which describe relativistic gravitation, can be derived by combining general thermodynamic considerations with the equivalence principle: he applies the relation $\delta Q = T\,dS$ to the local horizons of accelerated observers and uses the proportionality of entropy and horizon area [115]. This derivation can also be understood in terms of Landauer's principle: for the accelerated observer, matter that falls through the horizon is erased, and General Relativity amounts to the bound being saturated [116]. Within quantum field theory, the thermodynamic entropy can be replaced by the relative entropy that distinguishes coherent excitations from the vacuum; it is determined by the energy flux through the horizon, and if it is assumed to be proportional to the change of the horizon area, one obtains the semiclassical Einstein equations [117]. Erik Verlinde and others have investigated how alternative proposals for gravity might originate in entropic scenarios [118,119]. Jacob Bekenstein has also proposed a generalized second law: the ordinary entropy outside black holes plus the black-hole entropy never decreases [120]. Ted Jacobson later linked the Einstein equation to vacuum entanglement: for first-order variations of the local vacuum state of conformal quantum fields, the vacuum entanglement of small geodesic balls is stationary if and only if the Einstein equation holds [121]. In a program aimed at deriving gravity from quantum mechanics and utilizing several mostly reasonable assumptions, the emergence of spacetime from entanglement, i.e., "bulk entanglement gravity", has been sketched [122].

If it is most difficult to reconcile gravity and quantum physics at the beginning of time it could be interesting to look at the other side, to the end of time. With expansion going on forever (let us assume), maybe even accelerating for some of the time, the universe turns basically into an empty void. Still, gravity might not peter out incessantly and drop to perfect (mathematical) zero. Time itself would freeze, and with it some bottom value of gravity. In addition, the very act of measuring would require some means, i.e., some type of apparatus, which cannot completely be devoid of mass or energy. This, in turn, would produce gravity, albeit very weak. At least some inevitably self-induced gravity thus will always be present for any observer [48]. This would be another analogy to thermodynamics, i.e., to the third law, saying that absolute zero cannot be achieved in finite time.

It is proposed to see gravity not smoothly dropping to zero in the weak limit as one argument for some type of modified Newton/Einstein gravity, MOND. Modified Newton gravity has recently booked some success in various versions; flat rotation curves of spiral galaxies [123,124] and, from weak lensing, indefinitely flat circular velocities of isolated galaxies [125], gravitationally weakly coupled binary star systems [126,127], the cosmic microwave background [128], and the observed bulk flow [129] can be described/explained. The result for wide binaries is disputed, however [130,131].

To account for gravitational lensing, earlier versions of MOND need to be generalized and relativistic. Maybe, in the end matters are more complicated, and some tensor-vector-scalar theory is required: the one proposed by Jacob Bekenstein [132,133] has been ruled out by the nearly simultaneous arrival of gravitational waves and light from the neutron-star merger GW170817 [134], whereas the newer theory of Skordis and Złośnik, which also adds a vector and a scalar field to the metric, lets gravitational waves travel at the speed of light and agrees with the observed CMB and matter power spectra on linear scales [128]. Einstein's General Relativity would suffer the same fate as Newton's gravity before: i.e., rendered a most appropriate and useful approximation in certain somewhat limited domains.

Probably it is all too early to dismiss Einstein or tinker too much with his equations. Taking the nonlinear field self-interaction effects of General Relativity seriously can possibly explain both, dark matter and dark energy [135,136]. This proposal is disputed: an explicit analysis concluded that confinement-type effects cannot play any significant part in explaining flat galactic rotation curves [137], to which [136] is a reply. Exploring the expansion of the universe using fundamental thermodynamical concepts for adiabatic conditions, some cooling has to be expected in a fluid approach [138]. This can be described with a Grüneisen-parameter, which is found as naturally embodied in the energy-momentum stress tensor in the Einstein field equations. A concept of "thermal time", with dynamical laws fully and only determined by correlations, has been considered as a possible basis for fully general-relativistic thermodynamics [139]. In the presence of gravity, temperature is not constant in equilibrium in space. This Tolman-Ehrenfest effect linking thermodynamics and gravity can be derived by applying the equivalence principle to a key feature, i.e., the "speed of time", which is the ratio between the flow of thermal time and the flow of proper time [62,140]. This speed of time in turn can be identified as the local temperature. Carlo Rovelli takes thermal time as centerpiece in an attempt to de-construct and afterwards re-construct time as the local order of events in a fundamentally relativistic account with time resulting from a blurred vision of macroscopic states, i.e., coarse graining [141].

## 8. No need for dark matter nor dark energy

As decades of dedicated search for dark matter (particles) and dark energy have so far turned back basically empty-handed, gravity as an emergent phenomenon might be worthwhile a consideration. Some version of MOND could deliver much of the observed performance ascribed to dark matter in galaxies (or better), although a residual missing mass of about a factor of two remains in galaxy clusters [142]; suitable approximations of General Relativity might do the same [135]. Heuristically, it has been sketched how space, gravity, and spacetime could emerge in a holographic scenario and rather directly yield Newton's law of gravitation [118,143]. Displacements change the entropy and lead to reaction forces. Gravity as an entropic force would then result from such changes in the information about the positions of material bodies. Generalization to relativity could further lead to the Einstein equations. The law of inertia might thus have an entropic origin following the equivalence principle. Entropic gravity can in a toy model also be linked to the quantum entanglement of small bits of spacetime information [144]. Any version of entropic gravity in which the entropy depends on the path a body takes faces an experimental objection, however: such records would destroy quantum interference in free fall, which is observed with neutrons [145] and with atoms [146]; on the basis of the neutron data, it has been argued that gravity cannot be an entropic force [147]. In the picture sketched here, free fall makes no records.

While it seems that there are several connections between quantum entanglement, gravity and dark matter, there might also be mechanisms for dark energy [148,149]. For the latter it might be even more natural to consider that our galaxy is located in a (not really very pronounced) under-dense region in the local universe. Such a local void has been proposed to explain the Hubble tension [150]; a void large enough to replace dark energy as a whole, however, is ruled out in its simplest (adiabatic) form by the kinetic Sunyaev–Zel'dovich effect [93]. In this context it is interesting to observe that MOND does predict more clumpy structures early on than plain (approximated) General Relativity / standard Lambda-cold dark matter (ΛCDM) models [129]. At the same time, early James Webb Space Telescope (JWST) observations of the young universe were reported to be in strong tension with ΛCDM cosmology, which seemed to favor hybrid models [151,152]. JWST observations yielded tensions with current models of galaxy evolution, detecting massive and apparently mature structures much earlier than expected [153,154]; other analyses have reduced this tension considerably [155,156], and the remaining excess can be explained by modest changes to the physics of star formation [156]. The hybrid model of [152] partly attributes the redshift to "tired light", whereas the time dilation of supernova light curves follows the expansion to about one per cent [76]. One particular way some issues with the Hubble tension and galaxy formation could be fixed is with an early contribution of dark energy [157].

In any case, adapting / enhancing low field gravity could look more promising than searching forever for seemingly directly non-observable dark matter or dark energy, the latter driving an observed accelerated expansion of the universe. Accepting General Relativity as basis, one should take its non-linear self-interaction fully into account. This seems to offer explanations for the effects ascribed to the presence of dark matter as well as dark energy [135,136]. While effective gravity would be boosted over shorter distances inside denser regions, its influence over wide distances in between massive blocks would be diminished. A clumpy universe can give rise to "timescapes", i.e., clocks ticking with different speeds depending on whether they are situated in widely empty space or inside massive dense regions [129,158]; the timescape model, which replaces dark energy by kinetic gravitational energy and its gradients, is reported to be very strongly favoured over ΛCDM by the full Pantheon+ supernova sample [159].

There might still be a little space for some dark baryonic matter in the vast expanses between far distant galaxies.

A large-scale asymmetry of the primordial fluctuations might explain the observed very small lopsidedness of the CMB [77] and an uneven distribution of very big structures observed today [129].

Based on the baryon acoustic oscillation measurements of the Dark Energy Spectroscopic Instrument (DESI), combined with the CMB and supernovae, dark energy has been hypothesized as waning ("thawing") compared to a few billion years ago [160]; in the second data release the preference over a cosmological constant is 2.8 to 4.2σ, depending on the supernova sample [161]. This could qualitatively fit with the non-linear self-effects of General Relativity, which would certainly not establish a single static value for a vacuum energy once and forever [135,136]. Dark energy according to the fluid model also turns out as time dependent in the current (dark energy-dominated) era [138].

## 9. No need or place for many worlds

It looks like all of the above can nicely fit with one real world, the universe, which we inhabit and observe. Quantum systems need classical framing while purported effects from un-reflected uses of unitary quantum physics seem dispensable and not contributing too much to our understanding. Nice and elegant as a theory, which needs nothing else but itself for its own interpretation, might be, it has been argued before that not only quantum but even classical systems always require some separate reference [162].

It seems clear that for physical models we have to stay inside the universe as observed (if we do not want to appeal to some metaphysical external god's eye perspective or help; doing this in effect by enthroning abstracted formalisms including the postulation of perfectly plain stability for some relations does not look like a convincing option).

Frames and records as suggested here, can naturally match with time as described in Special and General Relativity. It has been demonstrated that Einstein's equivalence principle can be generalized such that it applies for reference frames, which are associated to quantum systems, even when in a superposition of spacetimes [163]. This way, there seems to be no basic conflict between Quantum Theory and General Relativity. Nevertheless, that does not mean that arbitrarily massive bodies could be brought into spatial superpositions over large distances: phonons, which are inevitably excited when such a superposition is prepared, carry which-path information and put stringent constraints on it [164]. Superpositions have been demonstrated for molecules beyond 25 kDa [165], and Schrödinger cat states have been prepared for the vibration of a mechanical resonator with an effective mass of 16 micrograms [166].

Spacetime cannot be infinitely smooth. At some (very small) scale, fluctuations must appear due to the quantum nature of many observables, and also time [47].

Attempts to arrive at classical manifest observations and one solid world based on fleeting quantum mechanical behavior have basically followed two routes: decoherence (staying inside unitary QM formalism, but only for the entire system including the environment, and not for the respective quantum system alone) and modifying Schrödinger's equation (postulating external influences). Here, it is proposed to leave QM inside alone, but effectively delimit completely isolated quantum systems by relatively hard boundaries at their preparation (or birth) and also at collapse (possibly rather smooth at this end if partly isolated).

One mandatory ingredient required for reconciling the quantum with the classical realm obviously is a certain level of randomness; some type of fluctuations, as described by thermodynamics, thus play an important role in all accounts. Conceptualizing decoherence as transmitting / loss of information to an environment, end-boundaries are diluted as information leaks out of an incompletely isolated quantum system. With a tight relation between information and energy (in communication systems, the entropy production is at least equal to the information transmitted, a specific version of Landauer's principle [167]), it may be conjectured that in the end at least the amount of entropy given by Landauer's principle is produced for a quantum system to behave classically. Like heat from mechanical friction, information related to entanglement does not return back to resurrect some exact starting conditions; this is the arrow of time.

The standard von Neumann entropy is not suitable for characterizing extractable work from internally correlated systems, any lack of knowledge limits the amount of work, which an observer can gain [168]. Strict energy conservation and the Jarzynski fluctuation theorem cannot be observed at the same time when extracting work from a quantum system in a thermal state [169].

In a very recent proposal of marrying quantum mechanics and gravity, fluctuations feature prominently, and it is hypothesized that small masses can be measured as fluctuating [170,171]. Zero-point fluctuations with particle-antiparticle pairs leading to a polarizability of the vacuum also play a decisive role in an attempt to address the cosmological enigma and derive (some) dark energy as vacuum energy from Casimir self-interaction of quantum electrodynamic fields [172]. Local contributions from Casimir self-interaction would most probably not preclude effects due to field self-interaction in General Relativity over wide distances [135,136], and the other way round.

## 10. Other parallels to the outside of the physical box

Any complete understanding of the universe has to cope with unknown unknowns and surely with principled limitations. These refer to the sheer understandability of physics on the one hand side, and to limited capabilities of humans on the other. The "unreasonable effectiveness of mathematics in the natural sciences" as described by Eugene Wigner [173] does not come with a guarantee for infinite extension. On the contrary, mathematical impossibilities surely also constrain physical models. Additionally, any structure and theory must somehow match with human capacities for perception and cognition. Paraphrasing Immanuel Kant: "the conditions of the possibilities to experience objects are at the same time the conditions for the possible objects of this experience" [174]. Technical instruments built on the basis of thorough physical understanding have dramatically expanded human perception, often leading to serendipitous discoveries. In some (not so far) future, similar might happen to constrained human cognition. The Ouroboros Model offers an explanation how our subjective experience of unfolding time structured in moments arises in iterations [1,174].

Looking at biology, we have learned that nature is a tinkerer, and assuming one divine streamlined master-plan behind all physical reality might simply be misguided; and if nature were that clear in some end, that structure would only be accessible to us via cognition, which in turn is principally limited and relying to a large part on manifold abstractions, which are linked and interwoven, but rather specific for clearly demarked contexts [1].

The "natural" but not naïve interpretation of quantum mechanics described above has to be considered local; everything we really know of happens in a (classical) context. Any prima facie non-local effect can solely be detected by employing *meta*- selections the specifications of which itself cannot be transmitted without relying on at least some classical communication. Quantum mechanics is thus left in a situation akin to the case of mathematics: the impossibility to demonstrate some basic tenets of quantum reality without resorting to classical means can be seen as corresponding to Gödel's incompleteness result. Some "external" reference for non-contradictory self-consistent grounding is required for "completeness". This fundamental open-ended nature of relevant endeavors (in mathematics and for physics) does not prevent beautiful and useful results, on the contrary.

David Hilbert's program of strict and complete axiomatization of mathematics (and physics) has been proven impossible. Kurt Gödel's incompleteness proof, just like Alan Turing's work, has self-referring statements at its core. The demonstrated irresolvable contradictions can be ascribed to a clash between a sought-for "eternal" (timeless) mathematical solid structure and a "dynamical" twist to it, which contraposes different "moments" (i.e., contexts). If that famous guy from Crete had made reference to the respective actual time in the real world and exclaimed "I lie now: (in this following interval, with this specific well-defined statement)" nobody would have ever bothered much.

In the light of the above, taking time fully into account looks like offering the best way out, and this often offers a resolution of apparently infinite regress. Bhartṛhari and Julian Roberts have had that idea (long) before [175].

Anyway, there are (individual) limits to human cognition and understanding. Natural laws carved out as straight(ened) important links in the web of dependencies have to be "accessible and simple", and this constraint can only be expanded so much with artificial intelligence. Relations are selected as fundamental if they are clear and direct, useful, compatible with (almost) everything known, "elegant",…; i.e., if they are in a way "beautiful", similarly as in daily life or art. It can be admitted that such a program might not be simply "realistic", as there will be always something which we do not and cannot know, but the ambition and the obtained results overall certainly are not "anti-realistic".

The topic of complexity with a main distinction between problem classes belonging to P (solvable in polynomial time) versus NP (solutions checkable in polynomial time; for the hardest of them, the NP-complete problems, no polynomial-time algorithm is known) might have some additional bearing on solidly establishing (and limiting) a basis for our (and machine) understanding of the world, which is principally achievable. A very rough analogy between graphs and quantum systems can be seen by identifying nodes with (classical records of) events and links with (timeless quantum mechanical) developments between these specific points. There can be very many arrangements traversing each link exactly once, while there is a limited number of possibilities when nodes (events) are only encountered once on each route; in the complete graph with five nodes, for instance, there are 132 Eulerian circuits but only 12 Hamiltonian cycles, although the numbers depend on the graph.

The difference between the two problems is demonstrated in the nice animation in [176]. While only slow exponential algorithms are known to find a Hamiltonian path, any route can be checked for correctness in polynomial time. For Euler graphs, both finding and verifying correct routes is easy with a fast polynomial algorithm. In Quantum Mechanics, events (with irreversible real records in entropy) are fewer than possible connections between them (links are manifold and timeless).

It is tempting to speculate, whether some type of Feynman integral approach, taking all possibilities into account, could be adapted and whether an associated Schrödinger-type equation for NP-hard problems could be found. Quantum computers might then have an edge while classic simulations implemented with neuronal networks would be quite efficient up to some capacity limits. Roughly having all hubs and links in the mind at the same time and then starting with the strongest, i.e., internally most densely connected, clusters (coarse graining) in an iterative procedure is hypothesized to be a procedure by which humans address such problems; e.g., the one called travelling salesman. Intermittently fixed anchoring points pave the way to a solution and fend off capacity limitations to some extent (at the expense of time required for a solution) [177].

Foregoing absolute optima helps to find near-optimum solutions efficiently. Bayesian processes as recently often proposed as blueprint for cognition, are NP-hard [178]. Limiting the material to be considered in a given context, is one way to make it tractable [174]. In the end, the reachable universe is finite; this establishes some overall hard boundary conditions.

**11. Conclusions**

In a down to earth perspective, one specific result and immediate practical consequence of the above should be mentioned. Taking seriously that the transition of a system between quantum and real has a threshold given by Landauer's principle, it matters that this threshold is an entropy, k ln2 per bit, independent of temperature; temperature only sets the energy price, kT ln2. A hotter environment holds more excitations that can carry information away and therefore usually decoheres a system faster; above a critical temperature, entanglement even vanishes completely (see below). Coherence is protected by weak coupling together with level spacings well above kT of the environment: optical photons stay coherent at room temperature, while superconducting qubits near 5 GHz must be cooled to millikelvin. When trying to harness quantum "weirdness", e.g., for quantum computers, one should thus focus on degrees of freedom whose energy quanta are large compared with the thermal energy of their surroundings.

On the meta-level, which is in the focus of this article, emancipating ourselves and renouncing the reliance on some external clockmaker or unique eternal formalism (and their respective perspectives), we have to assemble our understanding of the universe from what is rationally accessible to us inside. This follows from an obvious criterion of intellectual honesty. While that attitude does not exclude the contingent existence of other spheres [67], these cannot have a decisive place in any truly scientific account. The widest possible consistency between all known different constituents and their strongest achievable and testable interlinking is all, which we can aim to reach at a certain point in time. Approximating the emergence of reality by quantum decoherence seems to deliver similarly with basically the same accumulation of entropy in the environment in the end. Since noise renders every state mixed, one may ask whether among all states with a given spectrum there is a maximally entangled one; such states do not exist for every spectrum [179]. Above a certain temperature, independent of the size of the system, Gibbs states of local Hamiltonians are separable, i.e., entanglement vanishes completely; this has been called a "sudden death of thermal entanglement" [180], a term that echoes the finite-time "sudden death of entanglement" of noisy qubits [181]. The here emphasized "collapse" would be just a special "abrupt" case of the general process. There are limits to the application of the formalism, it can even be argued that in a somewhat paradoxical twist objects could hypothetically decohere to profoundly non-quantum superpositions of massive bodies [164,182].

So, some type of "Ouroboros"-arrangements appears to be the best, which we can ever achieve. For the here advocated three-legged approach the Triskelion might be a fitting symbol, see Figure 1. If one insists, in a very abstract sense, this symbol could be taken as an encompassing God's-eye view of the physical universe.

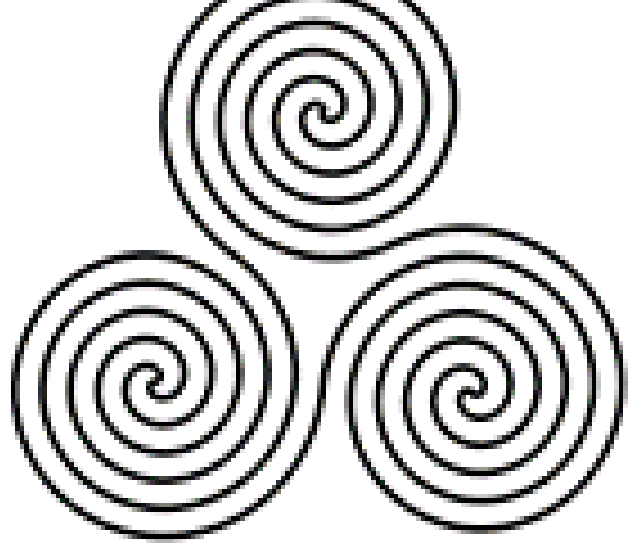

**Figure 1.** Triskelion symbol as an abstract shorthand for how Quantum Physics, Thermodynamics, and Relativity are interlinked and mutually induce each other [183].

For the whole ideograph all three legs are necessary, no one is more basic or important than the others. No vicious circle nor indiscriminate associationism is meant but an overarching consistent narrative with well-defined building blocks consistently embedded in some meaningful (time-) structure. Circles and loops are perfect, but more is required than just making ends meet; spirals (and helices) are preferred and they have to have large enough diameters (and heights), embracing the knitwork of the entire universe, at least potentially as far as can be seen at any given time. In this somewhat fleeting picture strongly emphasizing emergences, one might try to find some basis starting with a concept of energy.

The situation actually is the same, for physical theories, especially about the beginning of the universe, as it is for understanding consciousness and Free Will [174]. Also there, self-reference causes no problems as long as time is taken into account properly. Staying inside the known and rationally/scientifically accessible universe, widest-ranging and well-organized self-referral is the best one can hope for. This entails pushing boundaries.

The proposal here is to see space as primary basis for a better overview and put grainy emerging time as “main culprit” for unavoidable fluctuations, which are essential for linking the quantum and real domains, at the forefront. Time itself is anchored in the expanding space of the universe, all void without material content [52], at least as far as its direction is concerned; the rates of clocks are set by local physics. Fluctuations themselves are “timeless” and not really “real” as long as they do not produce entropy and records, just as in isolated quantum systems.

The important point then is that releasing gravity from standing a solitary pillar to some emergent status and tying it closer to thermodynamics could somewhat close a braided picture just the same as for the case of time. A self-consistent and intricate interplay of mutually self-reinforcing dependencies like between quantum and classical mechanics appears to be the best that we can strive for. This is quite similar to the relation between spacetime and gravity where neither space nor time would exist without energy or massive objects, the latter also experiencing gravity in the spacetime warped by masses, which in turn sense accelerations in time.

Identifying quantum physics as basis for classical phenomena while quantum effects becoming real only when suitably framed by classical events, then only seems appropriate. Similarly, understanding space and time as emergent from an underlying microscopic substrate, dovetails with real irreversible time being established only in interactions and always involving some type of records and classical entropy generation. It is no deficiency that the highlighted derivations of General Relativity leverage ideas, which had been developed from observations and models of gravity before [115,118,119]; one has to iteratively use the material, which is available (at the time in question). The point is that the overall picture is grounded, coherent, and consistent with all available, in particular experimental, evidence.

What applies to dark matter, i.e., giving up a fruitless search for whatever exotic particle and rather accepting a view of emergence in the real universe might be applicable the same for dark energy. While an increasing production of vacuum energy appears promising by attributing some repulsion to the emerging number of possibilities with growing space [172], living in a relatively empty local bubble might also be worthwhile a consideration, at least for the Hubble tension [129], and so might the timescape model [159]. The proposal of non-linear self-interaction in General Relativity appears to offer a promising avenue for research as it could even account for changes (reductions) in dark energy over longer time scales as recently reported [135,136,160,161]. This conspires with new findings when modelling the universe as adiabatic fluid [138].

Running against the fundamental human wish for simple explanations, there is no reason why nature should not have settled with a constellation combining many diverse effects and contributions, on the contrary.

A compound conceptualization with growth plates ("Wachstumsfugen", the German word is better by highlighting the malleable space between more solid parts) instead of smooth and seamless uniformity might be the best way to describe our many-faceted universe. The overall picture is more like a mosaic than a smooth and continuous canvas. In restricted specific contexts (in praxis, with different approximations and associated ranges of applicability) if --> then causal dependencies can be sought for and often successfully delineated. Different perspectives corresponding to different models and approximations can be most useful and effective in some contexts but detrimental for addressing other problems [135, 136]. In an effective web of concepts and relations, more than one path between any two points can be expected. Actually, the denser the web and the finer the mesh the more complete is the picture, the deeper and encompassing the understanding, which then effectively unites many diverse perspectives [174]. This does not fundamentally exclude an underlying "plan of a watchmaker"; actually, nothing inside the universe could do this, just the same as for confirming one.

The author hopes that with this very rough sketch a thicket of intricate formalisms could be recast in an emancipated, enlightened, and "democratic" perspective, and the whole somewhat ordered in a self-consistent, coarsely systematic, at least not completely haphazard way (although some measure of uncertainty or chaos appears indispensable in order to do justice to a full model of the universe). The "three pillars" in the end are abstractions themselves in a web of manifold dependencies and (abstractable) regularities. Bending trajectories to straight as suggested with the Alena Tensor, offers unexpected new perspectives and links [184]. "Time" in this picture appears as a most versatile throughgoing abstraction related to the order of events.

There is nothing like "substance", which would obey the definition of Baruch de Spinoza, i.e., something that is "in itself and is conceived through itself, that is, that whose concept does not require the concept of another thing, from which it must be formed" [185]. Recent investigations in quantum reference frames lead to similar conclusions regarding localization and time leaving no ground for "absoluteness" [7]. The concept of substance like that of physical laws are abstractions, i.e., Platonic figures, grounded in, but detached from many details of a braided underlying fabric of reality. Reality is everything together, emerging from the interplay of many diverse strands and relations. Unquestionably reachable from the inside, there is nothing like one fundamental particle, or one Theory of Everything, or one formalism, or one basic eternal truth, and neither any one God. There is nothing like an Archimedean point.

Facts are not relative/private other than described by Relativity and inherent limitations for communication, even if there can be many diverse perspectives. A detailed formalism allows for the full reconciliation of relativistic and quantum notions of causality. Experiments performed in classical spacetime can be explained in terms of a definite and acyclic causal order at a fine-grained level [64,65,66].

Ultimately, overall consistency with all accessible boundary conditions at a time and including developments is the touch stone for "truth", i.e., fully convincing models, which allow to explain known facts and to formulate interesting predictions.

An ontology like the one by Nicolai Hartmann is demanded, latest as soon as (self-) interactions and combinations begin to build up higher-order phenomena [67]. While a certain reaction between two entities interacting might be the only possible one in a certain context, a lot of development and time might have been necessary in order to arrive at that particular state. Gestation of an embryo and a full bird inside an egg need very little input in terms of external energy/heat but the process starts from a most peculiar point with a tremendous history to draw effective information from. Synergetics as developed by Hermann Haken and emphasizing time offers the best chances to describe how different levels of organization can interact and make novel features and structures emerge [186]. Hermann Haken and Philip Warren Anderson realized amongst others the importance of sheer quantity of any constituent [187]. The very concept of emergence can be traced at least to Nicolai Hartmann, and probably much longer back [67]. Long ignored, it recently received some interest with the advent of new mathematical formalisms with causation going bottom up as well as top down [188,189,190,191].

Following this line of thinking, it is tried here to sketch that a self-organizing approach with mutual dependencies, constraints and promotions, leading to the unfolding emergence of our one reality, can form a suitable basis for the existence of the universe and our understanding of it. Starting from any one chosen "cornerstone", the others can be found, explained, and understood iteratively; in the triskelion symbol: tracing one uninterrupted line through its full length and through all turns.

The overall encompassing picture is claimed to be one of ("mutually induced") emergence. As "more is different" holds already in the context of quantity of one part alone, it should come as no surprise that this principle is valid when many diverse contributions are effective together [187]. The main focus of attention related to emergence usually is on attempts to explain some higher level of organization on the basis of the involved constituents. In valid examples, the other side of the coin is obvious but often neglected; any overarching emerged structure will somewhat inevitably be undermined as soon as one delves into specific deep details. The grand big picture can easily get out of sight, and it might in fact even be destroyed by too limited myopia.

Transformations involving energy or information in the real world are inevitably associated with losses for a source and with entropy production. Some principal uncertainty alone and, especially, adhering to some type of energy/time (entropy/time) uncertainty relations, forbids 100 % efficiency and perfectly sharp demarcations. Even beyond that, Quantum Mechanics is not necessary for the unpredictability of the future of any sufficiently complex system. Entropy is not the same as information or a simple lack thereof. Entropy is relative and information is information only for a prepared (suitably knowledgeable) recipient; coarse graining prevails, and it is most often the case that only fractions of the total content can be transmitted and decoded. A most stupid guy, probably not even able to read a single word, knows that burning a (holy) book hurts the targeted audience. Collecting all photons emitted in the process could not bring back the meaningful content.

Even if still striving for some type of unification, a non-hierarchical lay-out of physical theory has to some extent been proposed before by Carl Friedrich von Weizsäcker in the format of a "Kreisgang" (walking in a circle) through a web of relations and dependencies [192]. By establishing records at disruptions, the unitary evolution of quantum states, which is information-preserving and time-reversible, can be reconciled with the global evolution of the universe following the second law of thermodynamics, which, in general, is neither. Poincaré recurrence is a mathematical construct not applying to the real non-conservative world; even inside the quantum realm, ergodicity can be broken because of destructive interferences [193].

Emphasizing a braided lay-out in a dynamic "process-view", self-referring and with high interconnectedness, could be seen as an attempt of combining Eastern and Western traditions, which, amongst others, have repeatedly been found to put different weights on context as part of the respective cultures [194].

To what extent the widespread search for one theory of everything (like earlier: the philosopher's stone) can be traced to a preference for monocausal thinking (probably grounded in monotheism or following from the same roots) would be another interesting topic; – for history, cognitive science, and sociology. “Beauty” and “elegance” ascribed to (mathematical and physical) theories (as well as to pieces of art) are hypothesized to follow from the same roots in human cognitive processes [1,174].

Letting go of any form of unique metaphysical goal-directedness, the here suggested lay-out and path forward is probably but one of several or many possibilities to approach an overall consistent picture (each emphasizing different constituents and relations). This should be seen as encouraging and as genuine witness of possibly achieving some comprehensive understanding covering many facets and including various diverse chains of arguments.

(Physical) theories (conceptualizations, models) are not all equal; they can self-reflectively and -consistently be ranked according to criteria including how big their “diameter” is, i.e., whether or not they cover a large range extending over many layers in an ontological hierarchy, how widely applicable and accurate their results are, how solid they appear, how deeply grounded and based on well-established facts without leaving large gaps, how important their field of application actually is, whether they are open to progress or better even promote improvement and growth, and many more; – with some of the demanded attributes and their weights most likely changing (mostly slightly) with the accumulation of sound knowledge over time. Identified grounding layers with many (causal) links emanating are in a privileged position. Uncertainty relations / trade-offs appear to be essential as one can never be sure to have taken all potentially relevant parameters in interesting contexts into account properly, – except in very restricted cases.

For a very first shot aiming at a really big scientifically fully grounded and coherent picture without appealing to supranatural powers, the author took the liberty to suppress many details leaving a lot of room for serendipity. Attempting a coarse but encompassing view, how it all could fit together, hopefully helps to turn attention and effort to promising directions [174].

The above sketch deliberately leaves one fundamental point open: the loose ends are bundled in the question of how it all began, in a very special state of low entropy. This is, on the one hand, about the best one can hope for in any attempt to draw a truly encompassing picture and, on the other hand, the minimum required to guard the proposal against accusations of promoting some dogma, one horn of Hans Albert’s Münchhausen trilemma [195].

**Note added in v4:** Several statements of earlier versions have been corrected: the cosmic expansion provides a common cosmic time and, in the view taken here, the direction of time, but not the rates of local clocks; the entropy of the final state of an ever-expanding universe does not return to zero, only the entropy density goes to zero; the cosmological redshift of free-streaming radiation is adiabatic and produces no entropy; the universe has had no centre at least since nucleosynthesis, while its state close to the beginning remains open; a common slowing of all clocks only relabels time, so that a “slow genesis” requires a change in the ratio between different clocks; the threshold set by Landauer’s principle is an entropy, guaranteed per cycle rather than per act of measurement, so that high temperatures do not favour quantum devices; the agreement of different clocks does not result from synchronization; and the trade-off between precision and dissipation holds for stochastic clocks, not for all clocks. Further statements have been updated in the light of recent literature, and the references have been checked. The agreement of clocks and the role of records as grains of time are worked out in more detail in a paper in preparation.

**Acknowledgments:** Very insightful and supportive comments by Časlav Brukner on [13] via private email are gratefully acknowledged. Special thanks go to some colleagues for their inspiring comments. This version is a careful revision of v3, carried out with the support of Claude, an AI language model by Anthropic, which checked statements and references against the cited sources and proposed corrections and additions. All changes were reviewed and decided by the author, who takes full responsibility for the content.